\documentclass[12pt]{iopart}
\usepackage{iopams} 
\usepackage{graphicx}
\begin{document}

\title[RELAXATION OF STATES IN DISTRIBUTIONS OF QUANTUM DOTS]{RELAXATION OF FERROELECTRIC STATES IN 2D DISTRIBUTIONS OF QUANTUM DOTS: EELS SIMULATION }
\author{C. M. CORT\'ES$^1$, L. MEZA-MONTES$^1$, R. E. MOCTEZUMA$^{2}$ and \hbox{J. L. CARRILLO$^1$}}

\address{$^1$ Instituto de F\'{\i}sica, Benem\'erita Universidad Aut\'onoma de Puebla, \\Apartado Postal J-48, Puebla 72570, Mexico}
\address{$^2$ Instituto de F\'{\i}sica de la Universidad Aut\'onoma de San Luis Potos\'i. Mexico}
\ead{lilia@ifuap.buap.mx}

\begin{abstract}
The relaxation time of collective electronic states in a 2D distribution of quantum dots is investigated theoretically by simulating EELS experiments. From the numerical calculation of the probability of energy loss of an electron beam, traveling parallel to the distribution, it is possible to estimate the damping time of ferroelectric-like states.  We generate this collective response of the distribution by introducing a mean field interaction among the quantum dots, and then, the model is extended incorporating effects of long-range correlations through a Bragg-Williams approximation. The behavior of the dielectric function, the energy loss function, and the relaxation time of ferroelectric-like states is then investigated as a function of the temperature of the distribution and the damping constant of the electronic states in the single quantum dots. The robustness of the trends and tendencies of our results indicate that this scheme of analysis can guide experimentalists to develop tailored quantum dots distributions for specific applications.
\end{abstract}

\noindent{\it Keywords}: Quantum dots distributions; Ferroelectric relaxation time; Bragg-Williams approximation; Electron energy loss spectroscopy
%\maketitle

% ================================================================

\section{Introduction}
\label{sec:intro}

The knowledge and possible manipulation of the electronic relaxation time have been a subject of fundamental relevance in many areas of condensed matter physics, quantum optics and information technology. The decay times of quantum states is closely related to the lost of quantum coherence whose control may have important applications in the field of quantum computation, particularly in the preservation of quantum information  coherence in a noisy environment. 
In this sense, the exploration of the ultrafast phenomena that determine the %c
relaxation times even though they %
are longer than the coherence time, has become a fundamental issue in these topics of research.
In this work we focus our attention in the possibility of changing the decay time of collective states in interacting quantum dots (QDs) by changing their characteristic parameters and those of their 2D distribution. The generation of entangled states in quantum dot systems, the observation of their decoherence in times of the order of a  picosecond as well as the transfer of coherence from a quantum macroscopic state have been  demonstrated several years ago \cite{reyesa,rodriguezf,potz,bhwu}. Presently it is possible to design these nano-scale systems that behave as artificial atoms, in the sense that they posses a well defined and controllable spectrum, at least at certain extent \cite{billaud,krenner}.\\
Since many of the physical properties and possible applications of these quantum systems depend on the eigenstates of the confined  electrons \cite{martab,axell,bimberg}, by controlling the physical characteristics of the interactive QDs of a distribution, one is able to change the decay time of the collective states. The later one would depend on a very complex way on the eigenstates of the single quantum dot, as well as on several physical characteristics of the distribution. This is a highly relevant question that this work attempts to contribute to answer, by means of a simple but physically meaningful approach, based on well established approximations. We suggest Electron Energy Loss Spectroscopy (EELS) as a proof of concept. The insights reached on this phenomenology would provide some tools and possibilities to learn how to control the coherent lifetime of the electronic states in quantum dots, that has been largely a clue issue to increase the feasibility of using them to develop some important  quantum devices \cite{zrenner}. Then, the understanding of the influence of the electronic properties of the single constituents on the emergent collective properties of these systems is truly important. \\
The collective  behavior of the QDs has been experimentally studied by means of EELS \cite{howiea}. This technique is used to analyze the behavior of the energy gaps of individual quantum dots as a function of particle size \cite{erni}. Furthermore, electron scattering in the Low-Energy version allows to analyze compositional aspects of quantum dots  \cite{beltranm}.
It has shown to be a really useful experimental tool to investigate the dipolar response of nanocomposites \cite{pitarke}. The whole dipolar response and ferroelectric behavior of QD distributions has been investigated theoretically, by means of the simulation of EELS experiments \cite{rosariom}. 
The only experiment related to our theoretical approach was reported in \cite{fuchs}, although it was applied to nanocomposites. \\
In this work the decay time of ferroelectric states of 2D distributions of self-assembled GaAs quantum dots, embedded in an $Al_xGaAs_{1-x}$ matrix is studied theoretically. The potential that confines the electrons inside the quantum dots is modeled by a step barrier potential approximation, and the inner and the outer regions have also different effective masses. The transitions between the individual quantum states  result from the interaction of the confined electrons with the electron beam, where the excited states decay after a life time described by a damping parameter ${\gamma}$.  The collective relaxation times are determined through the curves of the so-called energy loss function, as a function of the energy lost by the electron beam, for different values of $\gamma$ and temperature.\\
The interaction needed to generate a ferroelectric behavior in a distribution of QDs is introduced here by the inclusion of a wetting layer in the system \cite{melnik}, and in the respective Hamiltonian by means a Weiss-like mean field approximation. Additionally, in order to explore the role of correlations in the collective behavior of QDs, we analyze their effect by means of the Bragg-Williams approximation. This  theory has been applied to analyze order-disorder in grown quantum dots \cite{malachias} as well as in the calculation of magnetic properties of metallic nanoparticles \cite{chen}.   \\
The practical procedure is as follows. 
We solve the Schr\"odinger equation and calculate the eigenvalues and the eigenfunctions of a 2D distribution of conical self-assembled quantum dots. In terms of the calculated quantum states of the distribution we determine the electric susceptibility and from its behavior we obtain the critical temperature ($T_c$), at this point the system present a phase transition and is accompanied by symmetry breaking. Finally, we obtain the collective states due to the interaction with the electron beam, the response to external changes produces transitions and further decay later.  We infer the decay time of the collective states through the width of the probability of energy loss distribution. We can understand our thermal state is entangled for a given temperature, and their coherence time is related with the relaxation times of the collective system. 

% ================================================================

\section{Single Quantum Dots}
\label{sec:np}

We start our numerical procedure by computing the energy states and corresponding wave functions of conical QDs of GaAs with radii $15$ nm and $4$ nm, which are embedded in an $Al_{x}GaAs_{1-x}$ box of $45$ nm width. Unless otherwise is specified, 15 nm dot radius is considered. 
Geometry details are important for the determination of eigenstates and in turn the dielectric response \cite{arezky}. Different shapes have been considered in previous calculations. We have chosen the conical confinement potential and those sizes, because they lead to clearer results for the dielectric response and the susceptibility, thus a sharper estimation of critical temperature can be obtained \cite{rosario2d}. The conical shape presents only azimuthal symmetry, which is easier to break by means of an applied electric field. These QDs are therefore more easily polarized. As a consequence, properties near the ferroelectric transition are better estimated as mentioned before. New techniques have made possible the growth of nearly conical-shaped QDs connected by a layer \cite{cedric}
We solve numerically  the time independent Schr\"{o}dinger equation, considering the one-band envelope function formalism and the Hamiltonians are described below. The effective mass of the electron in the matrix is taken to be $m_e = 0.0962$, while  $m_e = 0.063$ in the GaAs QD. The potential step between the QD and the matrix is $V_0 = 500$ meV. The wetting layer is $2.5$ nm width. The external electric field is 1 meV/nm. We assume a realistic surface density of QDs of $1.5 \times {10^5 / cm^{2}}$. 

% ================================================================

\section{Collective States}
\label{sec:fb}
For several decades, quantum entanglement has been extensively studied, both with relation to the quantum mechanics foundations as well as to important applications \cite{schloss}.  We have studied the cooperative behavior of a system composed by dipoles pertaining to the quantum states of the QDs, by an approximation based on the Ising model. This  approach allows to reproduce the physics of phase transitions, which occur when a small change in a parameter such as temperature ($T$) causes a large-scale qualitative change in the state of the system. At sufficiently low temperatures, there is not much random motion, and the configuration of the dipoles lined up with the external field is highly favored, while at sufficiently high temperatures, the random thermal motion destroys much of the effect of the field \cite{rosariom}.\\
Fluctuations in the polarization are determinant in the behavior of the system in the critical region. Usually fluctuations are sensed by means of the correlation function which is given by $h(r_i , r_j)$. Typically these influences between the components of the system described with the spatial and temporal correlations, occur over a characteristic distance $\xi$, known as the correlation length.  A short $\xi$ means that distant dipoles are very weakly correlated. Scale correlation length is obtained from the correlation function $h(r)$ which depends usually on the spatial separation {$r$} between a pair of dipoles  $p_i$ and $p_j$. We take into acount only spatial correlations. Thus \cite{yeomans}

\begin{equation}
h(r_{i}, r_{j})  =   \Big<( p_i - <p_i>) (p_j - <p_i >)\Big>,
\label{hrij}
\end{equation}
where $r_i$, is the position vector of site $i$ and $<>$ denotes a thermal average. If the system is translationally invariant $< p_i > = < p_j >$ and $h$ depends only on $r_i - r_j$

\begin{equation}
h(r_{i} - r_{j})  =  <  p_i p_j> - < p_i > < p_j >.
\label{hr}
\end{equation}

If the dipoles are independent this quantity should be zero. Away from the critical point the dipoles become uncorrelated as $r \rightarrow \infty$ and hence the correlation function decays to zero. From equation (\ref{hrij}), the correlations are measured among the fluctuations of the dipoles away from their mean values. Near the critical region, correlation function decays to zero exponentially with the distance between the dipoles as 
\begin{equation}
{h(r)} \sim r^{-\alpha}  {exp({-r / \xi})},
\label{eq:hsim}
\end{equation}

where $\alpha$ is a constant. This is usually the case for large $r$ near criticality. In the critical region long-range order develops in the system, $\xi$ becomes very large and the equation (\ref{eq:hsim}) breaks down.

Evidence from experiments and exactly soluble models shows that in these conditions $h(r)$ decays as a power law

\begin{equation}
h(r) \sim r^{-{d-2+\eta}},
\label{eq:hdecay}
\end{equation}

where $\eta$ is a critical exponent, and $d$ is the space dimensionality. It has been shown that the correlation length scales with the temperature according to \cite{yshi}

\begin{equation}
\xi = {{|T-T_c |}^{-\nu}},
\label{eq:scala}
\end{equation}

where $\nu$ is the critical exponent which describes the correlation length scaling \cite{kadanoff}. 

\subsection{ Mean Field and Long-Range Correlation Approximations }
\label{sec:mfapprox}
Theoretically, the simplest interaction which allows to reach a ferroelectric response in the distribution is a mean field approximation. This is better known as the Weiss approximation and is established through a well known Hamiltonian, that considers a potential felt by any QD due to the presence of their  nearest neighbors

\begin{equation}
{  H_{WEISS}=- \sum_{i}^{N} p_{i}E_{ext} +  \sum_{i}^{N} p_{i} E_{mol} = - \sum_{i}^{N} p_{i} E_{tot} },
\label{eq:weiss}
\end{equation}

where $\vec p_{i}$ is the { $i$}th dipolar moment, $ E_{mol}= \sum_{j}^{k} \frac{1}{2} J p_{j} $, $k$ is the coordination number and $J$ is the exchange integral. Therefore, the problem of many interacting QDs has been reduced to a problem of $N$ independent dots on which an effective field, $E_{tot}=E_{ext}+E_{mol}$ is acting.
By construction, fluctuations are neglected in this field. To calculate the ferroelectric response within this approximation one has to solve the eigenvalue problem of the stationary Schr\"odinger equation with Weiss Hamiltonian,  $H_{\tiny WEISS} \Psi(r) = E_{\tiny WEISS} \Psi(r)$, where the confined potential is implicitly included.\\\\
On the other hand, in the Bragg-Williams approximation the interaction among the QDs is introduced by defining a short-range correlation $\sigma$, and long-range correlation parameter $L$, both defined in terms of pairs of parallel or anti-parallel momenta. The Bragg-Williams procedure qualitatively incorporates fluctuations by the introduction of $\sigma$ y $L$ as follows.
For simplicity, for a 2D distribution it is assumed that the interaction energy $J$ between the nearest neighbors is constant, the pairs of dipoles are denoted by $(+ +)$, $(+ -)$ and $ (--)$, with the corresponding numbers of configurations $N_{++}$, $N_{+ -}$ and $N_{--}$. Thus the energy is
\begin{eqnarray}
  {E (p_i)} &=& {E(N_{+},N_{-}  ,N_{++},N_{+-} ,N_{--} )}\\
{}&=& - p_i E_{ext}(N_+ - N_{-}) - J (N_{++} - N_{+-} + N_{--}),
\label{eq:Energy}
\end{eqnarray}

Rearranging the dipoles, select  $+$, and since every dipole is connected with its neighbors $+$, we have
\begin{equation}
{ k N_+ = 2 N_{++} + N_{+-}},
\end{equation}

and similarly  with dipoles $-$, they are connected with their neighbors $-$
\begin{equation}
{  k N_- = 2 N_{--} + N_{+-}},
\end{equation}
where $N_+$ is the number of QDs with dipolar moment along the $y$ direction, $N_{++}$ is the number of nearest neighbor pairs of QDs with dipolar moment also along the $y$ direction and $N$ is the total number of QDs in the system \cite{huang}.\\
Within the range $[0,1]$, the parallel and antiparallel populations are: {$N_{+} / N$ } and {$N_{++} / {(N k / 2 }$ )}, considering that they are connected with the density of one dipole (global range), and with the dipoles correlation  (local range) respectively,
\begin{eqnarray}
 { N_{+}  \over N } &=& {1 \over 2} {(L + 1)}  ,  -1 \leq L \leq 1,\\
{ N_{++} \over ({N k /2 })} &=&  {1 \over 4} {(\sigma + 1)}  ,  -1 \leq \sigma \leq 1,
\label{eq:ndip}
\end{eqnarray}

In this way the Hamiltonian can be written in terms of the long-range correlation only, in the form known as the Bragg-Williams approximation
\begin{equation}
{{H_{BW} (L) \over N }} \approx-{J k \over 2}{ {L}^2 }-E L,
\label{eq:hwb}
\end{equation}
where $E$ is the external electric field. \\
Given a 2D QD distribution of dipoles, their geometrical characteristics are taken into account in the $L$ parameter according to equation (\ref{eq:hwb}). In our calculations we adopt the value $L>0$ to describe a ferroelectric response.
To incorporate the scaling hypothesis, we connect the equation (\ref{eq:hwb})  with the temperature assuming a power law, equation (\ref{eq:scala}):
\begin{equation}
{{H_{BW} (L) \over N }} \approx-{J k \over 2}{ {\xi}^2}-E L,
\label{eq:hwb2}
\end{equation}
$L$ is zero if the dipoles system is unorganized. Physically, $L$ is the polarization per QD due to the interaction with the external electric field. Thus, the relaxation times of the ferroelectric collective modes in the bidimensional distribution of QDs can be affected by changes in the surface density of QDs. 
In summary, long-range order means the non-vanishing of  the correlation when correlation length approaches infinity. 

% ================================================================

\section{Results and Discussion }
\label{sec:res&dis}

\subsection{Dielectric function}
\label{subsec:dielectricfunction}

We briefly describe the procedure by which we calculate the collective dipolar response of the 2D distribution of quantum dots.

% ================================================================
% ===== FIGURE  ===================================================
\begin{figure}[ht!]
\includegraphics[scale=0.28]{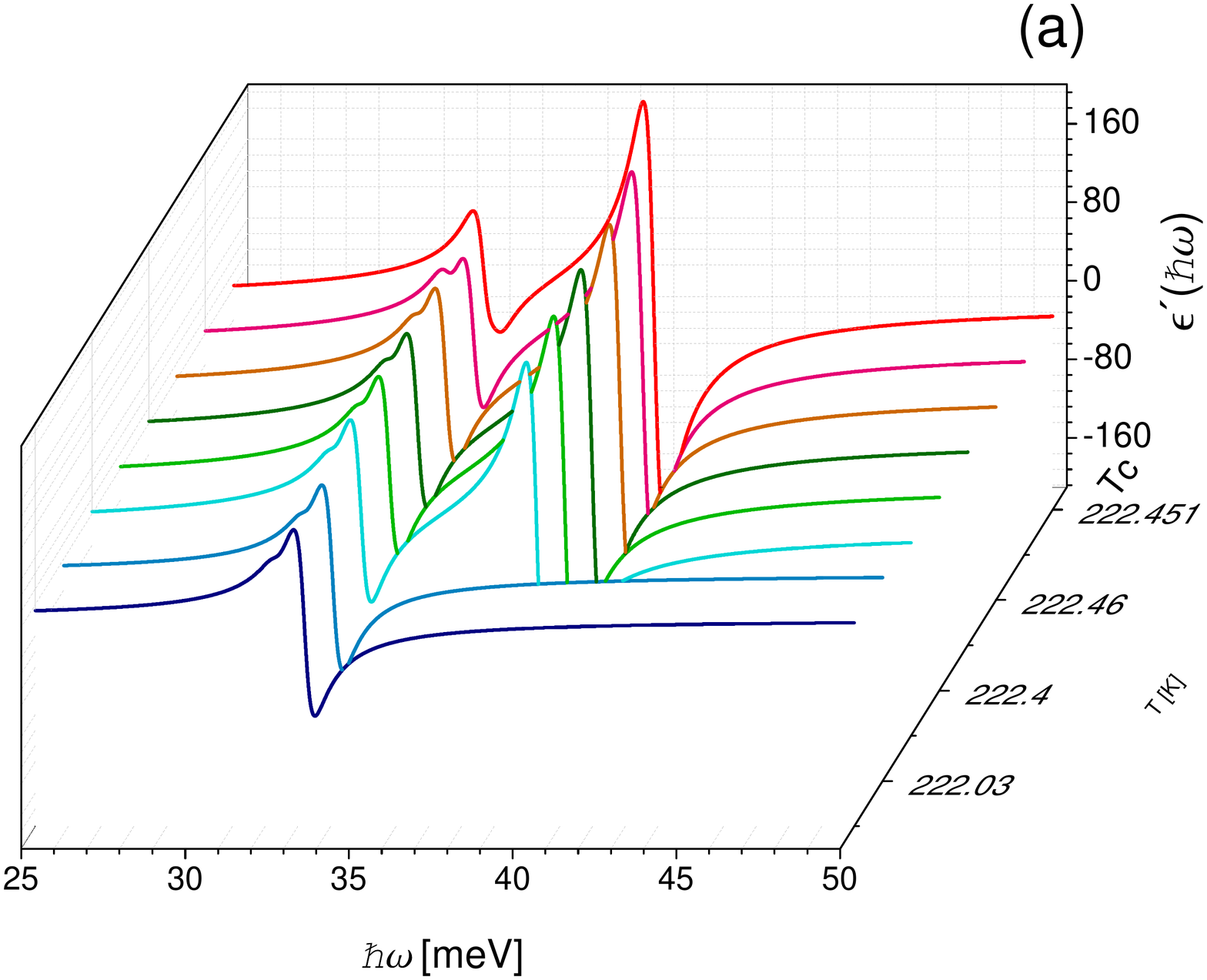}
\label{fig:im1}
\includegraphics[scale=0.28]{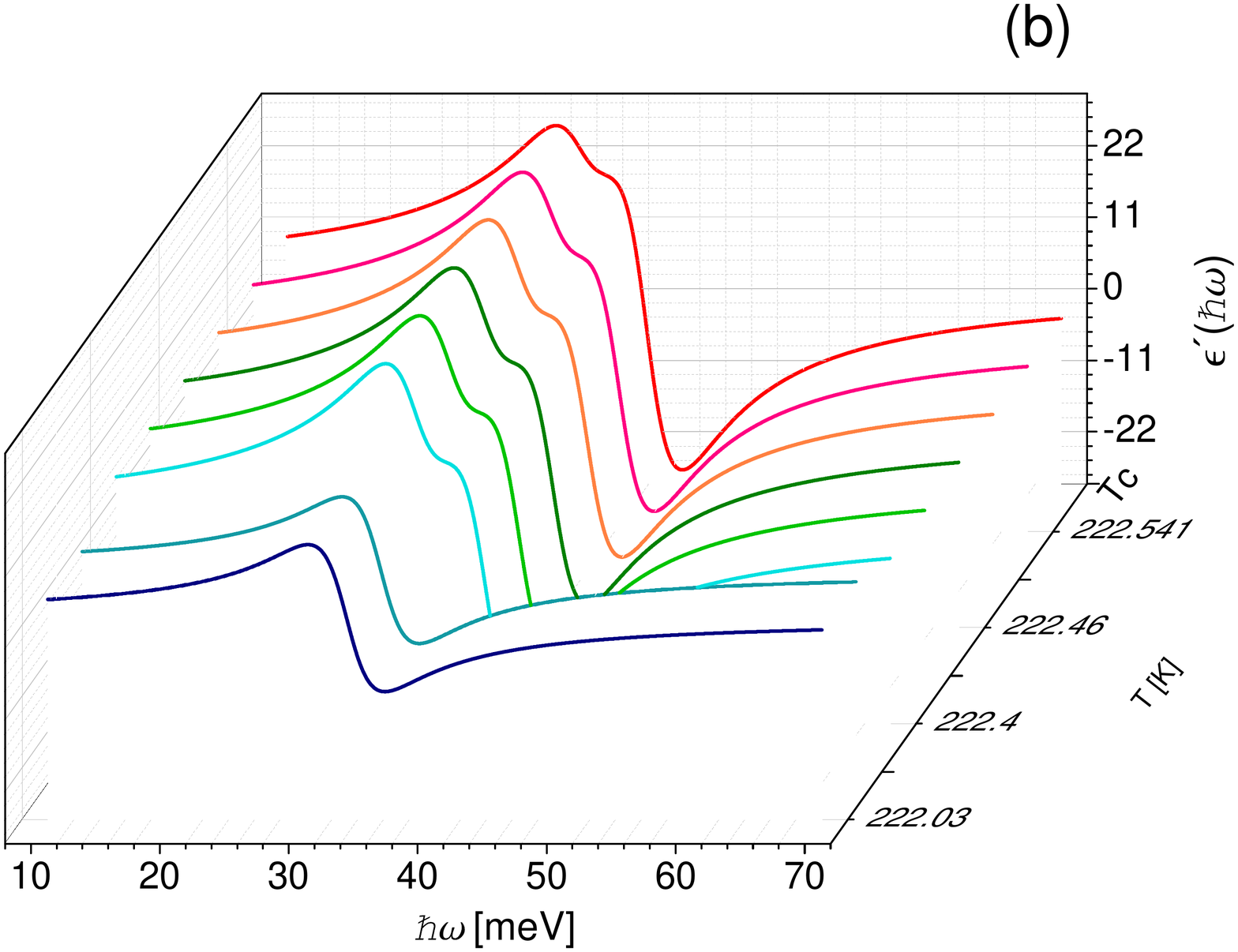}
\label{fig:im2}
\includegraphics[scale=0.28]{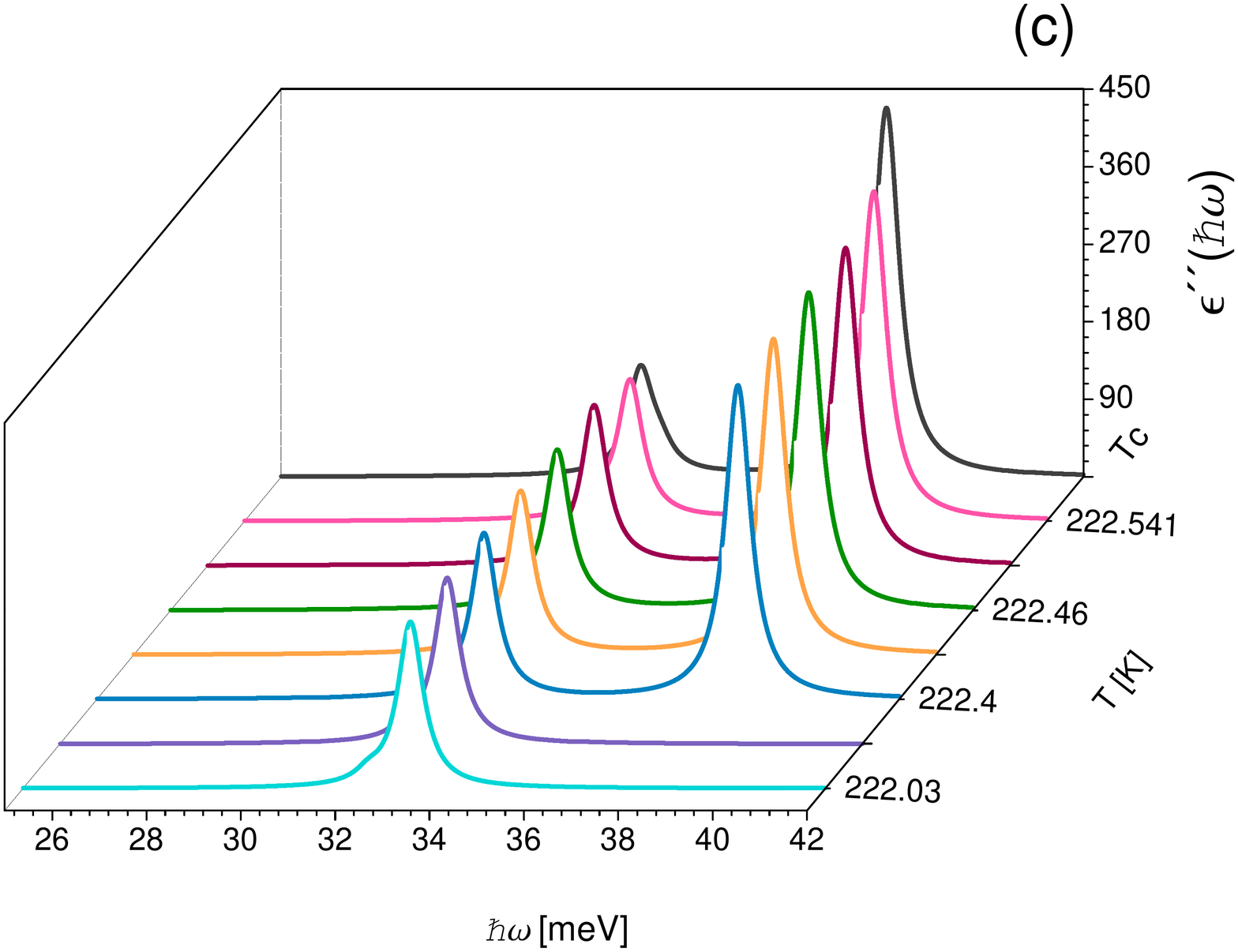}
\label{fig:im5}
\includegraphics[scale=0.28]{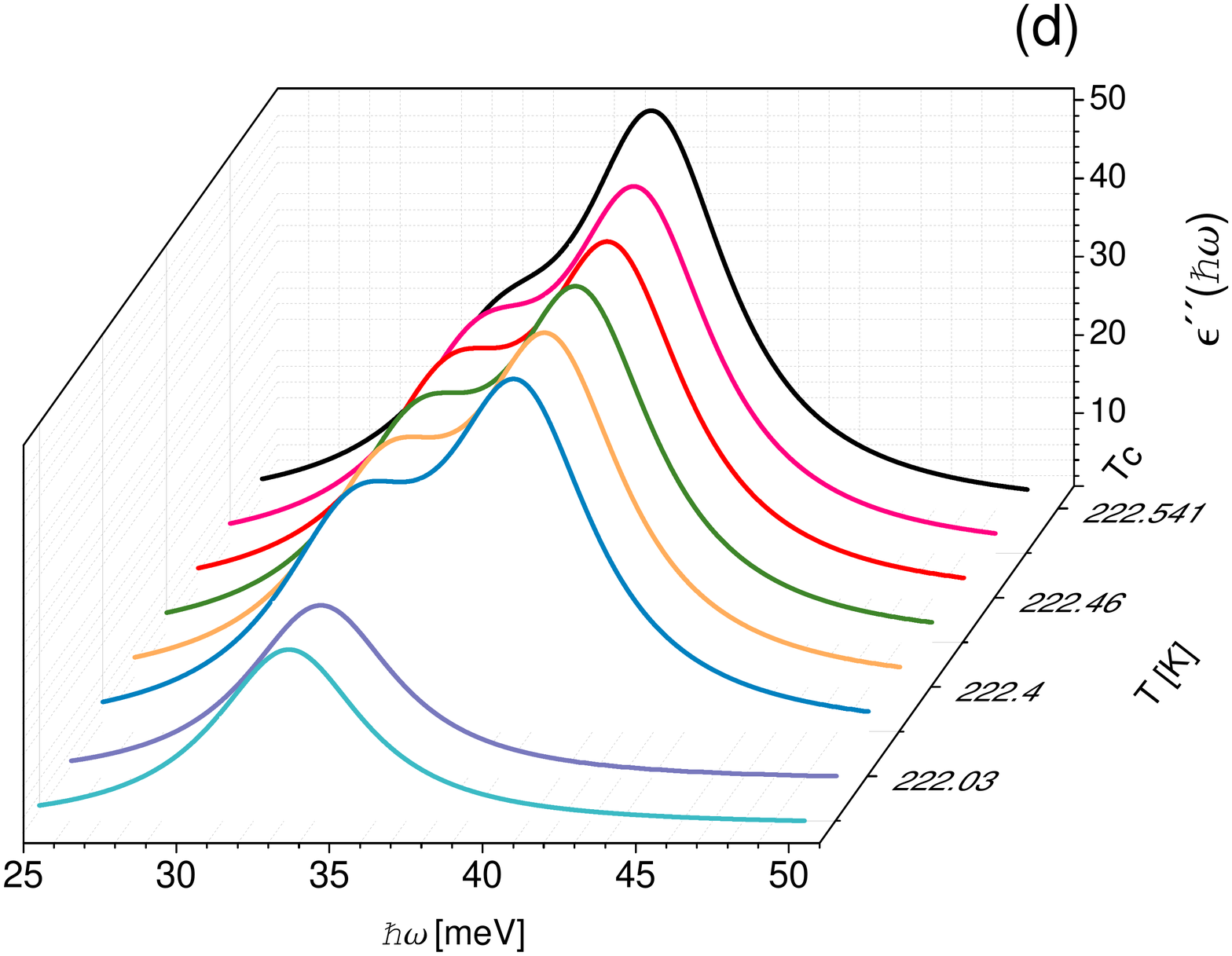}
\label{fig:im6}
\caption{The behavior of (a)-(b) real, and (c)-(d) imaginary part of the dielectric function as a function of frequency, for a distribution of conical QD in Bragg-Williams approximation. This approximation is able to predict three different resonance bands, although the widths depend on the damping parameter $\gamma$. (a)-(c) $\gamma = 500$ ${ns}^{-1}$, (b)-(d) $\gamma = 4500$ ${ns}^{-1}$.}   
\label{fig:edielectric}
\end{figure}
% ================================================================

We assume that the transitions between the quantum states of the distribution are caused by the interaction of QDs dipolar moments with the electric field produced by an electron beam traveling parallel to the distribution. Under these conditions it has been shown that the dielectric function of the distribution can be written as \cite{huag}
\begin{equation}
\epsilon(w) = 1- \frac{4 \pi i}{\hbar} \sum_{j}  |<\psi_{j} |x|\psi_{0}>| ^2 { \Big[\frac{1}{w+i\gamma-w_{j}} +  \frac{1}{w+i\gamma + w_{j}}\Big]},
\label{eq:epsilon}
\end{equation}

where $<\psi_{j} |x|\psi_{0}>$  is the matrix element of the transition between the zeroth and the {$j$}th state of the distribution, $\hbar\omega$ is the characteristic energy of the electrons of the incident beam, $\gamma$ is the damping coefficient and $w_j$ are the resonant or bonding frequencies. In semiconductor materials the most important dissipative mechanism that takes out energy from the electronic system in the quantum dots is the electron-longitudinal optical phonon interaction. The electrons may decay through longitudinal optical (LO) phonon emission The characteristics of the LO phonons in the QD system in a complex way strongly determines the decay time of the electronic quantum states. Features of the damping mechanisms like this are phenomenologically considered in the parameter $\gamma$. In order to get some insights about the excitation and relaxation processes in the distribution, we analyze  the behavior of the real and imaginary parts of $\epsilon(\omega) $. 

Fig. (\ref{fig:edielectric}) shows the real and imaginary parts of the dielectric function as a function of the energy loss of the electrons in the beam, for several values of the temperature. The Bragg-Williams Approximation with a single peak in the dielectric response function as observed at temperatures far from $T_{c}$, implies that thermal fluctuations are averaging out possible static deformation.
Notice the difference in the number of peaks, for different values of the lifetimes of the quantum states of the single quantum dots, described by the inverse of the  damping parameter $\gamma$. In the interval of frequencies and temperatures shown, it is clear that for the value $\gamma= 500$ ${ns}^{-1}$, a lifetime relatively large, the behavior of the real and imaginary parts of the dielectric function provides more information about the excitation and relaxation processes in the distribution. 
%For $\gamma= 4500$ 1/ns the real and imaginary parts of the dielectric overlaps.\\
By analyzing details like these in the dielectric function, or more precisely, in a function defined in terms of the dielectric function, it is possible to explore the influence of the characteristics of the QD distribution on the decay time of quantum states.

% ================================================================

\subsection{Energy Loss Function}
\label{subsec:energyloss}
In EELS, a high-energy electron beam travels parallel to the specimen to be analyzed.  Inelastic scattering of the beam  by the elementary excitations of the specimen changes the electron kinetic energy, and analysis of these energy and momentum exchanges provides information of the spectrum of elementary excitations. This technique gives also structural and chemical information of the sample \cite{Egerton}.\\
To investigate the collective quantum states of the distribution, we previously have proposed to simulate experiments of electron scattering in the EELS modality \cite{fuchs}. To do this we consider an electron beam with energy $ E_l = 100$ keV, traveling parallel to the surface that contains a distribution of conical non center-symmetrical QDs, and calculate the energy loss function $F(\hbar \omega)$, namely, the probability per unit path length and per unit energy interval of scattering with energy loss  $\hbar\omega$. For a 2D distribution of QDs this quantity is given by \cite{fuchs}
\begin{equation}
F(\hbar \omega)={\frac{K_{0}}{a_0 E_l}} \frac{2 z_0}{v_l / w}  Im\Big[{ {\epsilon_{M}(w)-1}\over {\epsilon_{M}(w)+1}}\Big],
\label{eq:fe}
\end{equation}

where  $K_0$ is the zero order Bessel function, $a_0 $ Bohr's radius, $E_l $ and  $v_l $ are the energy and velocity of the electrons respectively, and $z_0 = $ 1 nm is the impact parameter. 
Since $F(\hbar\omega)$ is a probability distribution, its behavior as a function of the energy loss is meaningful with relation to the relaxation times of the quantum states, it has an inverse relation with the average relaxation time ($\tau$) of these collective states. $\tau$ and the amplitude $\hbar\omega$ are related to the full width at half maximum (FWHM). \\
The loss function shows resonance peaks for certain energy loss values, resulting from the attachment of the field created by the beam of electrons with the dipolar moments of the distribution, where the quantum modes of the QDs are excited. 

% ================================================================
% ===== FIGURE  ===================================================
\begin{figure}[ht!]
\centering
\includegraphics[scale=0.26]{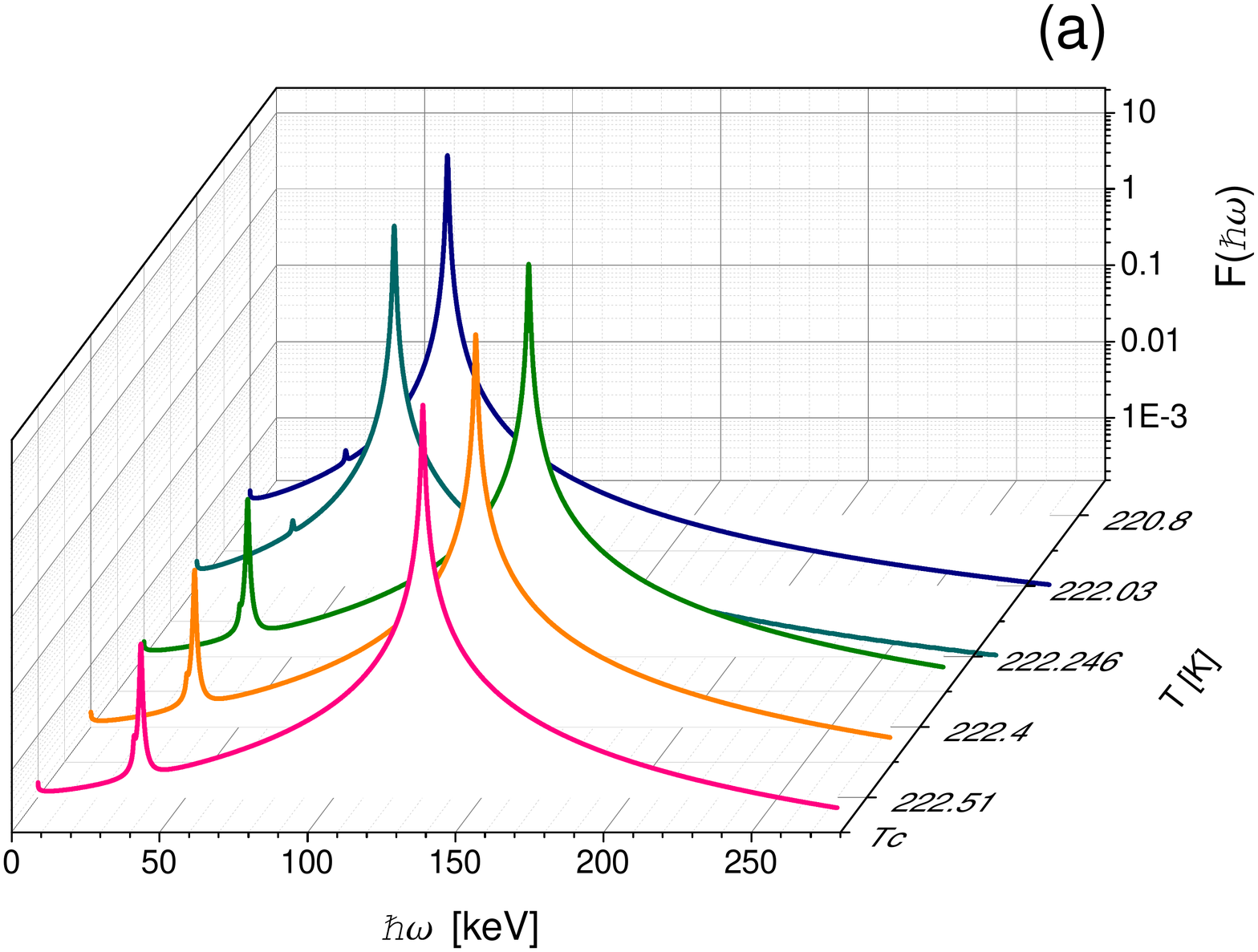}
\label{fig:fe1}
\includegraphics[scale=0.26]{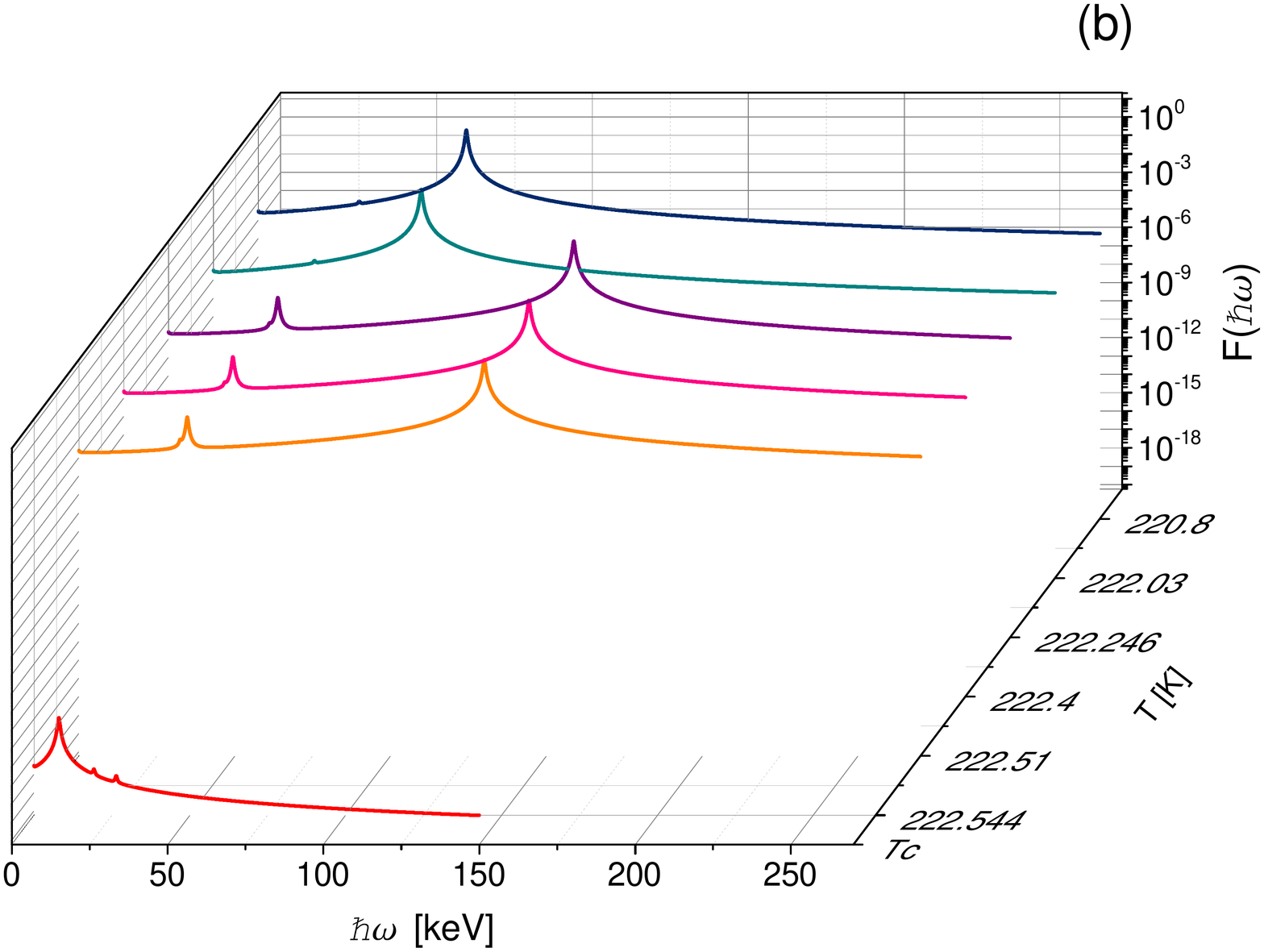}
\label{fig:fe2}
\includegraphics[scale=0.26]{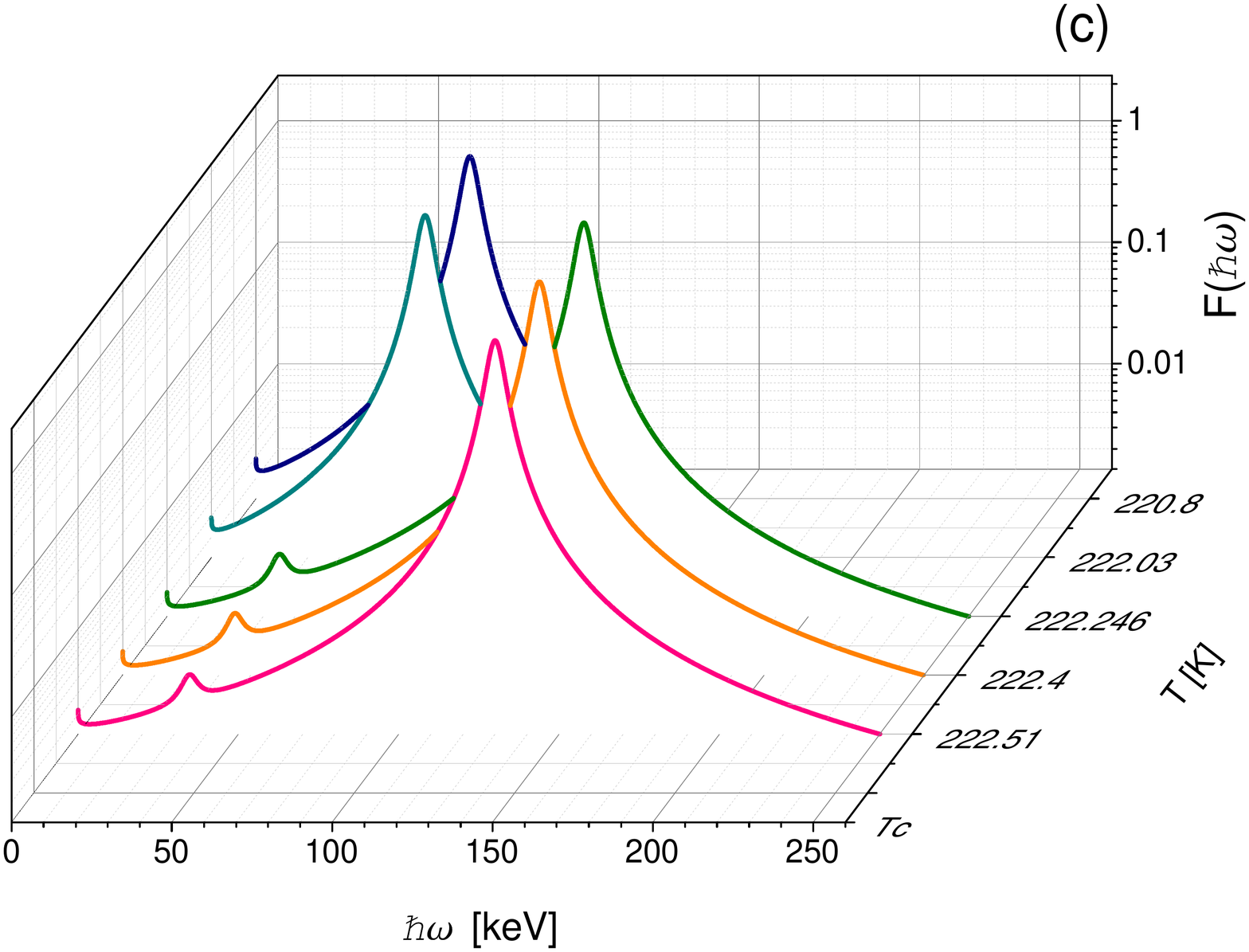}
\label{fig:fe5}
\includegraphics[scale=0.26]{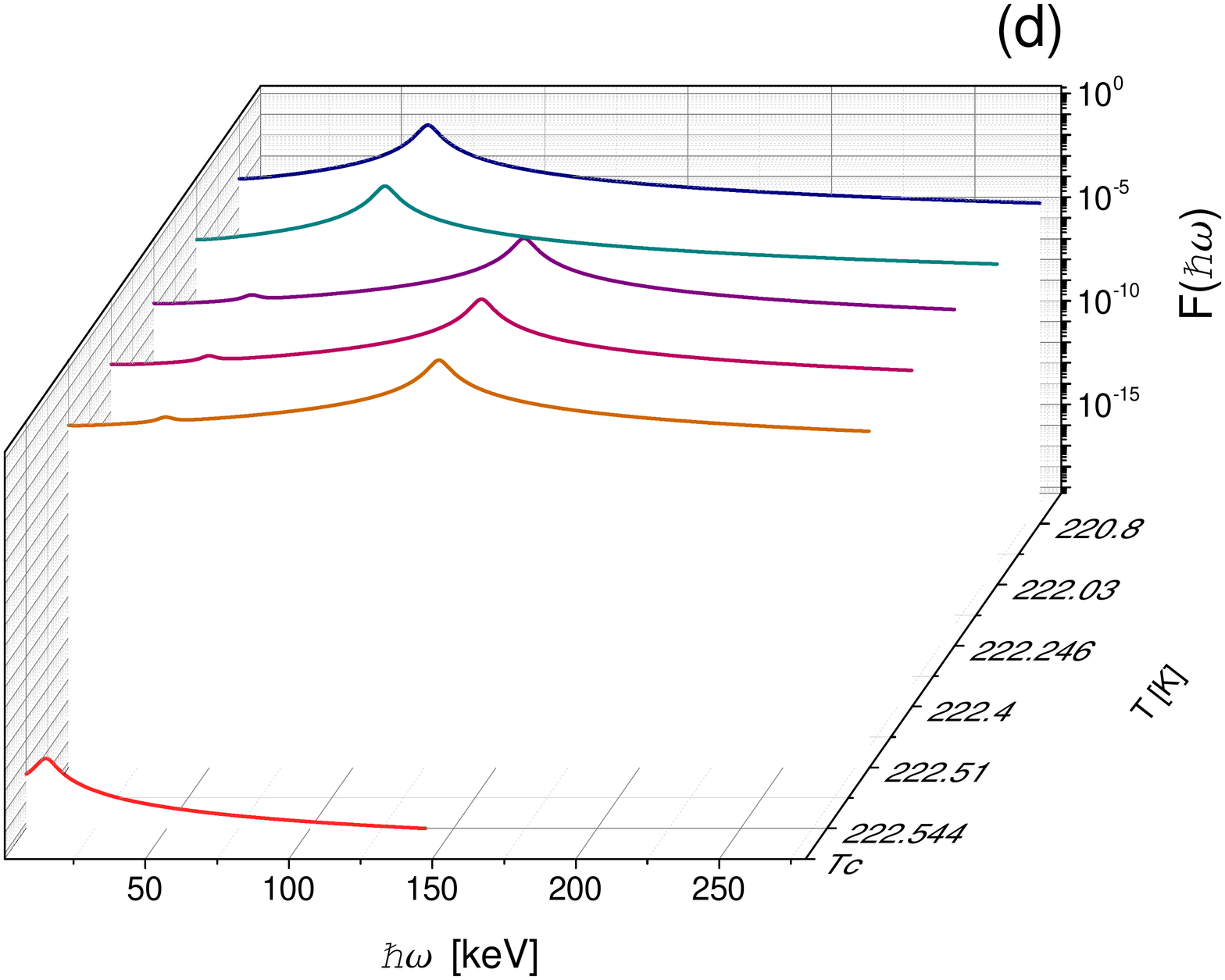}
\label{fig:fe6}
\caption{Energy Loss Function of a distribution of conical QDs with a 15 nm radius, in the Bragg-Williams Approximation. (a)-(b) $\gamma = 500$ ${ns}^{-1}$, (c)-(d) $\gamma = 4500$ ${ns}^{-1}$. b) and d) include the loss function at temperature near $T_c$ respectively. Notice the different scales.} 
\label{fig:feprobability}
\end{figure}
% ================================================================

In Fig. (\ref{fig:feprobability}) the behavior of $F(\hbar \omega)$ is shown as a function of the energy lost by the electrons in the beam, for two values of the damping constant, as predicted by the expression (\ref{eq:fe}). We have used the Bragg-Williams approximation to calculate the details of scattering processes as it was described above. Here the behavior of $F(\hbar \omega)$ again exhibits a better defined structure and higher loss probability for the value $\gamma = 500$. 
As temperature increases, correlation also does and higher energies are required to induce a resonance such that more energy is lost by the beam and the main peak is blue-shifted. Near $T_c = 222.55^o$ K in our case, dipoles are so strongly correlated that there is no exchange energy with the electron beam, such that loss probability diminishes, see Figs. \ref{fig:feprobability}(b) and (d).
The interpretation of the position of the bands of energy loss is completely consistent with that of the real and imaginary parts of the dielectric function, in the sense that the magnitude of the relaxation times of the quantum states of single quantum dots determine in a complex way the processes of relaxation and excitation of the collective quantum states in the distribution. The structure of the real and imaginary parts of the dielectric function and the structure exhibited by the energy loss function, reflect that through the coupling among the QDs, the system  generates collective states whose relaxation times depend on a complex way on the parameters of the distribution, but also on the parameters of the mean field, and extending this approximation to include fluctuations by means of the Bragg-Williams approximation, the structure of $\epsilon(\omega)$ and  $F(\hbar \omega)$ would also be affected. Thus, seeking at the effect that the characteristics of the distribution and the phonon distribution have on the damping parameter of the quantum states in the single QD, by the procedure here proposed one must be able to find general trends and features of the relaxation time of ferroelectric states of the QD distribution.

% ================================================================

\subsection{Relaxation Time}
\label{subsec:timerelaxation}

\begin{figure}
\centering
\includegraphics[scale=0.26]{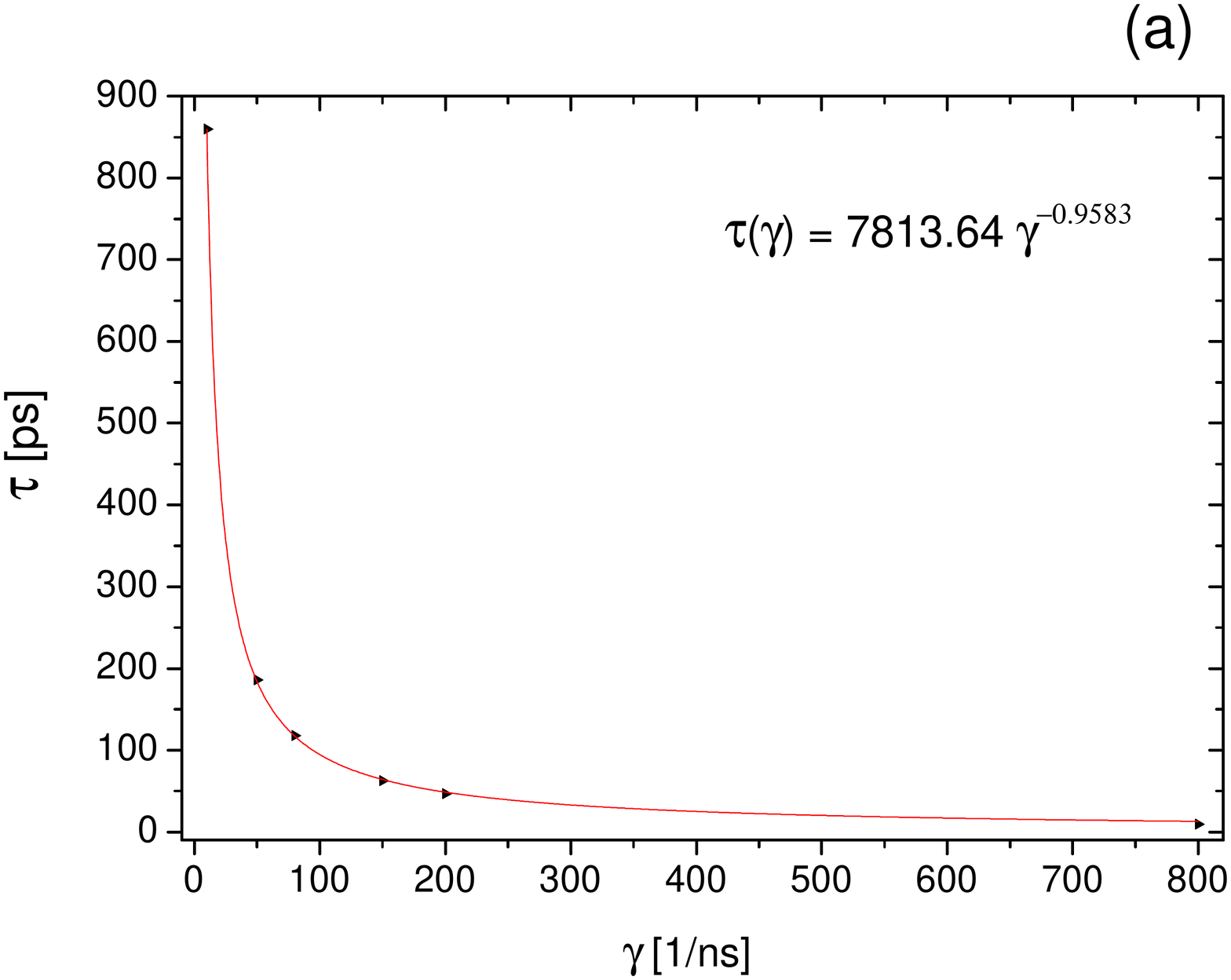}
\label{fig:t1}
\includegraphics[scale=0.26]{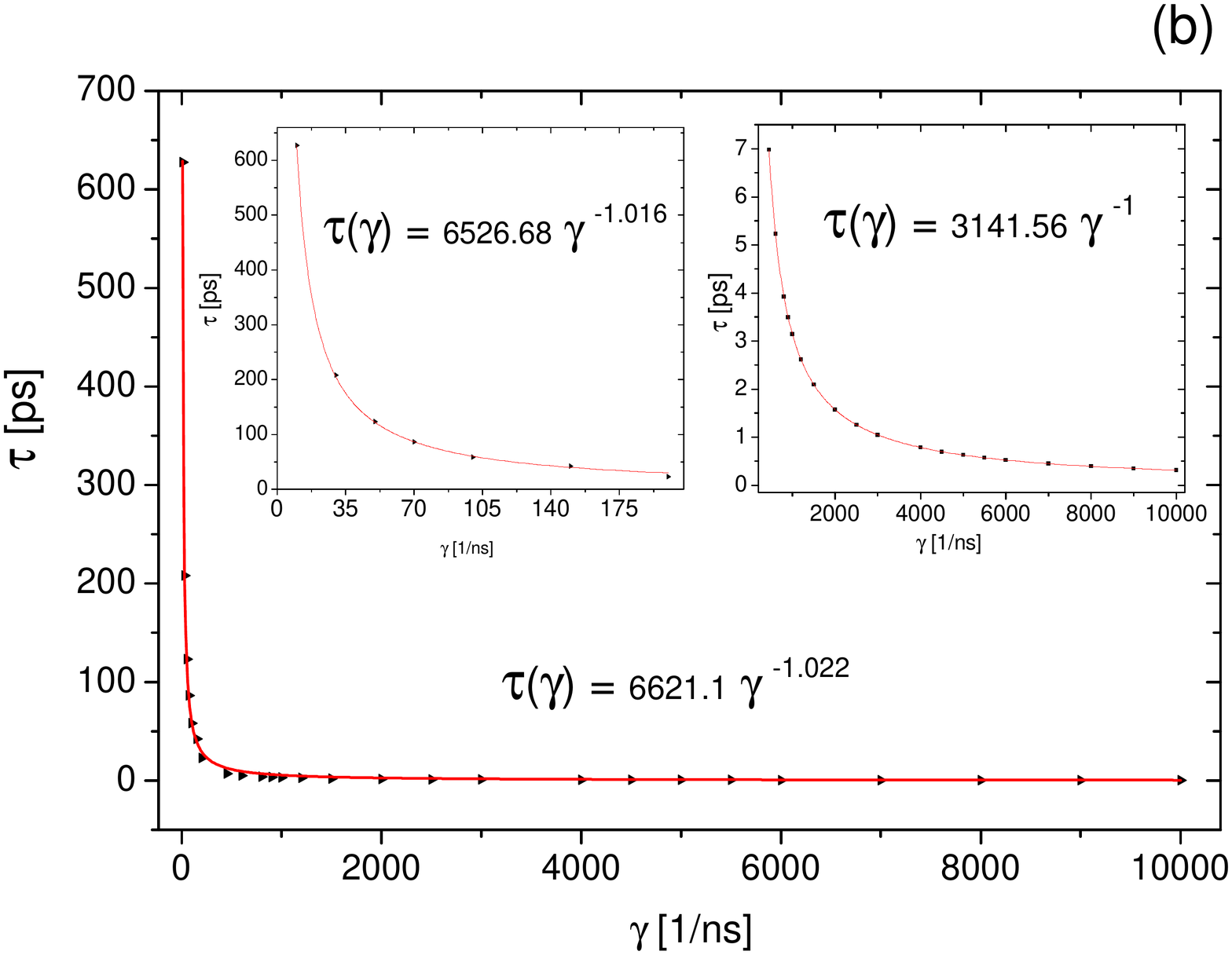}
\label{fig:t4}
\label{figuras}
\caption{Weiss Approximation. Relaxation times for different values of $\gamma$ of a distribution of conical QDs with radius (a) 4, and (b) 15 nm. Insets in b) show different ranges of $\gamma$. }
\label{fig:times}
\end{figure}

For lower values of $\gamma$, the quantum states excited by the electron beam will have longer lifetimes. Therefore, excited collective states will also remain longer. It is exhibited in sharper peaks in the loss function as Fig. (\ref{fig:feprobability}) shows. In the case of large $\gamma$, peak broadening yields in the $F(\hbar \omega)$ curves in such a way that the relaxation time of quantum states of different energies become convoluted. It is worth of mention differences with mean field (Weiss) approximation. To start with, relaxation times do not depend on temperature in this approximation and they exhibit a power law behavior with $\gamma$ as Fig. (\ref{fig:times}) shows, where the exponent has the value near 1. The relaxation time $\tau$ is larger for small dot radius. 

% ================================================================
% ===== FIGURE ====================================================
% ================================================================

\begin{figure}
\centering
\includegraphics[scale=0.26]{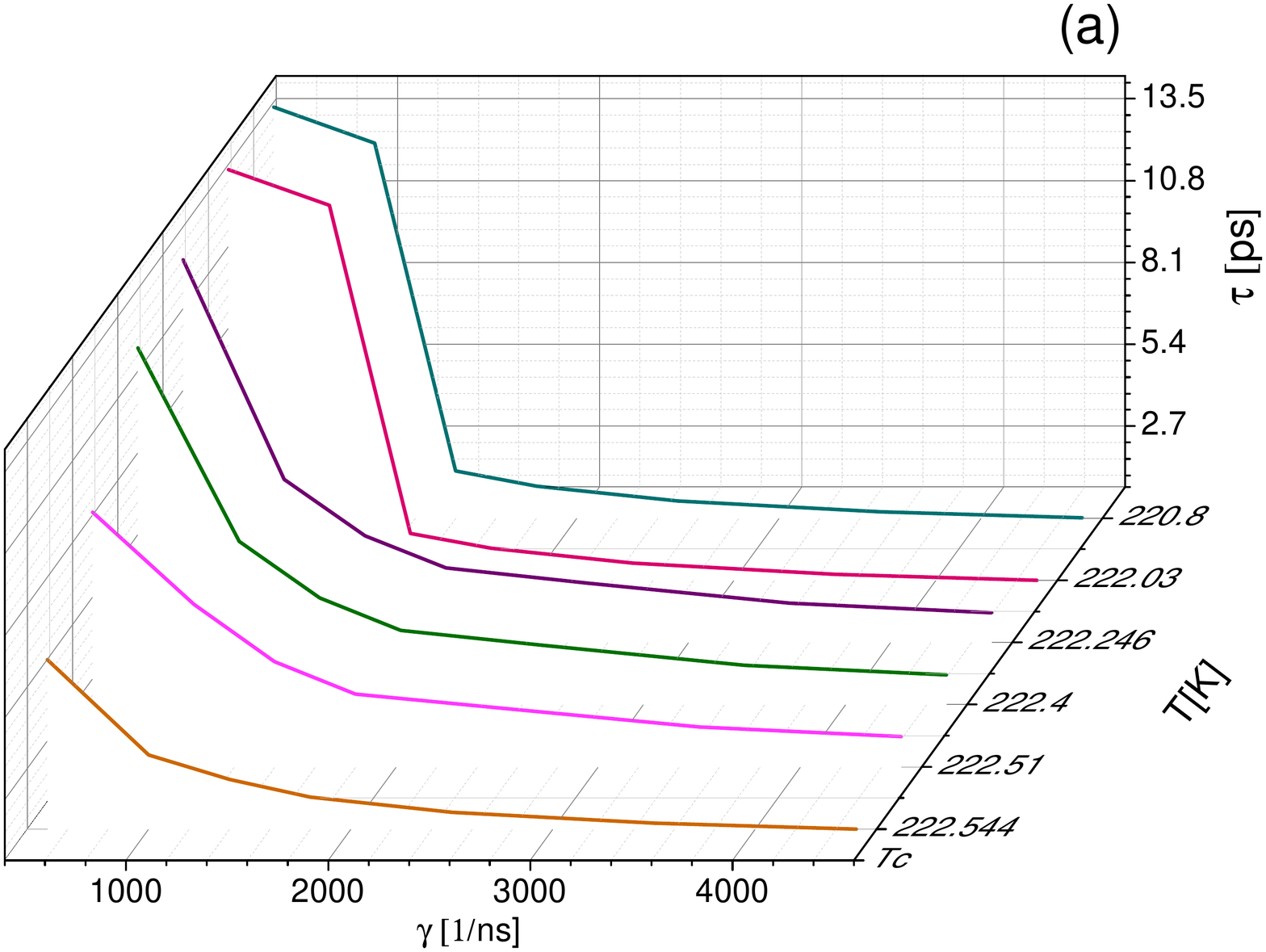}
\label{fig:t5}
\includegraphics[scale=0.26]{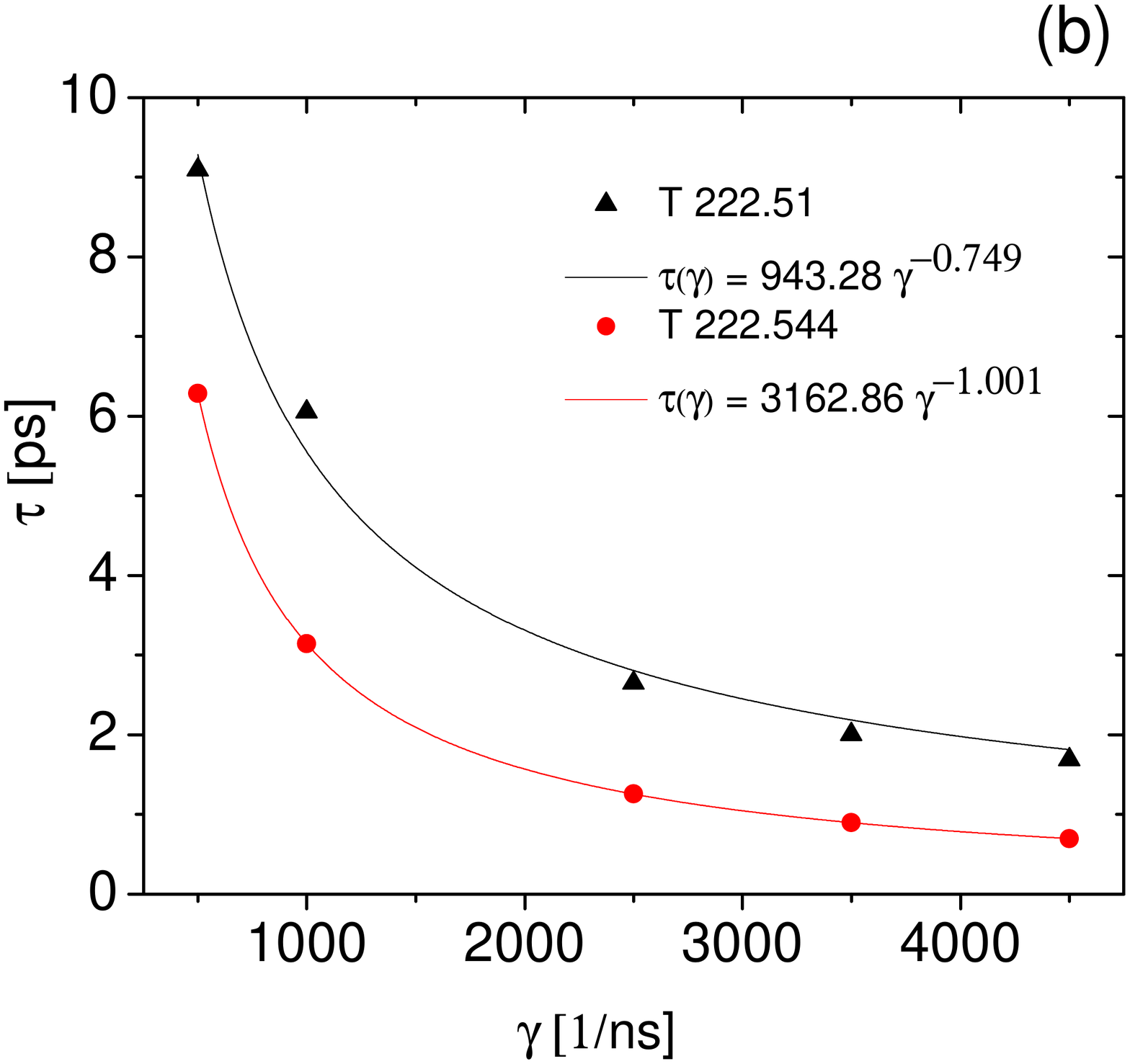}
\label{fig:t6}
\caption{Bragg-Williams Approximation. A distribution of conical QDs with a radius $15$ nm relaxation times for a) different values of $\gamma$ and temperature, b) two values of temperature showing different power-law dependence. }
\label{fig:t7}
\end{figure}

In contrast, the results provided by the Bragg-Williams approximation show a temperature dependence for the relaxation time as it can be observed in Fig. (\ref{fig:t7}).  We notice two regimes in the behavior of the relaxation time: At temperatures far from $T_c$ a complex dependence on $\gamma$ is observed. As we approach $T_c$, a power law decay appears whose exponent depends on the temperature, as observed in Fig. \ref{fig:t7}(b). Notice that in both approximations the  relaxation times are of the order of picoseconds.

% ================================================================
% ===== FIGURE ===================================================
%=================================================================

An increase of the width of loss function curves occurs when $\gamma$ increases, the second peak indicates that at relative higher temperatures an increasing number of collective states participates and the dipolar response of the distribution will also have a pronounced effect on the multipole response in the Bragg-Williams approximation. 
For instance, the dipolar response of short life time is characterized by the fragmentation of the distribution and spreading into the low-energy region, caused by non-resonant independent single-particle excitations. When $\gamma$  increases, the system has a short value of transition time between their states, so these dipoles have a weak interaction, and the resonances overlap. 

% ================================================================

\section{CONCLUDING REMARKS}
\label{end}

In this paper, we have studied the relaxation times of collective states of a  2D distribution of conical QDs. EELS experiments are numerically simulated for conditions near the ferroelectric-paraelectric  transition temperature. The loss energy function was calculated within the Weiss and Bragg-Williams  approximations. We found that the dipolar response is strongly sensitive to temperature, correlations and size of the dot. Relaxation times within the Weiss approximation do not depend on temperature and follow a power law dependence on the damping parameter of quantum states of single quantum dots $\gamma$, with an exponent nearly 1, which depends on the dot size. On the contrary, Bragg-Williams approach shows the existence of two regimes in the behavior of the relaxation time as a function of temperature. While far from the phase transition a complex behavior of $\tau$ is observed, near $T_c$ a power law describes well the observed behavior with exponents around 1. As expected, in the Weiss approximation a constant scaling law describes well the behavior of the relaxation time as a function of temperature, $\tau(T) \approx \tau^{-1}$. In the Bragg-Williams approximation two well differentiated power laws describe the behavior of the collective relaxation times at different temperature intervals. The robustness of  our calculations, based on the well established Weiss and Bragg-Williams approximations, indicates that this simple procedure could be a useful tool to  tailor QD distribution controlling at some extent the relaxation time of the collective states.  

\ack
Partial financial support by CONACyT and VIEP-BUAP. 
 
% ================================================================

\end{document}